\documentclass[12pt,a4]{article}
\usepackage{amssymb}
\usepackage[dvips]{graphicx,color}
\usepackage{wrapfig}
\usepackage{makeidx}

\makeatletter
  
  \@addtoreset{equation}{chapter}
\makeatother

\def\bfr{\begin{flushright}}
\def\efr{\end{flushright}}
\def\si{\quad}
\def\sii{\qquad}

\def\noi{\noindent}
\def\vs{\vspace}
\def\vss{\vspace{.5cm}}
\def\begc{\begin{center}}
\def\endc{\end{center}}
\def\be{\begin{eqnarray}}
\def\ee{\end{eqnarray}}
\def\eq{\label}
\def\nn{\nonumber  \\}
\def\rarwww#1{\smash{%
             \mathop{\hbox to 1.3cm{\rightarrowfill}}
             \limits^{#1}}}
\def\rarw{\rightarrow}

\def\d{\delta}

\def\e{\epsilon}

\def\l{\lambda}

\def\o{\o}
\def\p{\pi}

\def\t{\tau}

\def\f{\phi}
\def\F{\Phi}
\def\vf{\varphi}
\def\c{\chi}

\def\fr{\frac}
\def\del{\partial}

\def\mxxb{\left( \begin{array}{cc}}           %---- matrix 2x?
\def\mxxe{\end{array} \right)}
\def\mxxxb{\left( \begin{array}{ccc}}         %---- matrix 3x?
\def\mxxxe{\end{array} \right)}
\def\mxxxxb{\left( \begin{array}{cccc}}       %---- matrix 4x?
\def\mxxxxe{\end{array} \right)}
\def\mxxxxxb{\left( \begin{array}{ccccc}}     %---- matrix 5x?
\def\mxxxxxe{\end{array} \right)}
\def\mxxxxxxb{\left( \begin{array}{cccccc}}     %---- matrix 6x?
\def\mxxxxxxe{\end{array} \right)}
\def\mxxxxxxxb{\left( \begin{array}{ccccccc}}     %---- matrix 6x?
\def\mxxxxxxxe{\end{array} \right)}
\def\kakkob{\left\{ \begin{array}{c}}
\def\kakkoe{\end{array}  \right. }
\def\vecb{\left( \begin{array}{c}}
\def\vece{\end{array} \right) }

\def\bm#1{\mbox{\boldmath #1}}
\def\bm#1{\mbox{\boldmath #1}}
\def\bb1{\bm{$1$}}

\newfont{\bg}{cmr10 scaled\magstep3}
\def\deff{\stackrel{\rm {\small def}}{=}}

\def\pdel#1#2
{
\fr{\del {#1}}{\del {#2}}
}

\def\ppdel#1#2#3
{
\fr{\del^2 {#1}}{\del {#2}\del {#3}}
}

\begin{document}

\bfr
DIP 1901-01\vss
\efr

\begc
{\LARGE\bf  Another Counter-Example to Dirac's Conjecture}
\vs{1cm}

Takayuki Hori\footnote{email: hori@tokyo.zaq.jp}\vs{0.3cm}

Doyo-kai Institute of Physics, 3-46-4,
Kitanodai, \\
Hacniouji-shi, Tokyo 192-0913, Japan
\endc

 \vss

\begin{abstract}
Another counter-example to Dirac's Conjecture
is presented, which resembles the Cawley model but 
is so modified as to include second class constraints.
The arbitrary function in the general solution
to the defining equations of momenta
satisfies a non-linear differential equation.
Dirac's conjecture is examined for some solutions
to the equation.
\end{abstract}

\section{Introduction}

The problem of Dirac's conjecture has been
sometimes argued by various authors since
Dirac stated it\cite{dirac_1964}.
One group of them agrees that it holds, while
another group  does not.
The reason of the disagreement seems to be
in the setting of the conjecture and the
procedure leading to the canonical formalism.

The original statement by Dirac himself can be
expressed as 
that every transformation generated by first class constraint, which we call Dirac transformation, maps a state 
to its physically equivalent one.
According to an interpretation the above statement
is equivalent to that the appropriate hamiltonian
is the sum of the canonical hamiltonian and
linear combination of all first class constraints
including the secondary ones.
Sugano et al.\cite{sugano_1} argued that the above hamiltonian does not work in some models.
However,
in his canonical formalism,
there seems to be no rigorous proof of the equivalence
to the lagrangian formalism.
On the other hand  Deriglazov et al.\cite{deriglazov_1}  
showed that there is an extended lagrangian, which is obtained by adding auxiliary variables and is classically equivalent to the original lagrangian, where the Dirac transformation of the original lagrangian
is the physically equivalent transformation 
in the theory described by the extended lagrangian.

A system may be described by various lagrangians
which are classically equivalent but different from each other in auxiliary or
unphysical variables.
The constraint structures of the various
lagrangians are different and the validity 
of Dirac's conjecture depends on the form of
each lagrangian.
Indeed the gauge symmetry itself is an 
artifact emerging from 
the degrees of freedom of the unphysical variables.
Hence 
the meaning of Dirac's conjecture
in Deriglazovs' formalism seems to be different from that of original Dirac's one.

The author of the present paper has
proposed a canonical formalism of singular 
system\cite{hori_6} without extending the phase
space,
which is a straightforward generalization of that
of non-singular system.
The method is based on employing
the {\it general solution} for the velocity variables,
$u=\hat{U}(q,\p)$,
to the defining equations of momenta,
$\p=W\deff\del L/\del u$.
The procedure of the method is logically
clear compared with that of Lagrange multiplier
by Dirac. The latter is mysterious according to Deriglazov\cite{deriglazov_2}.

For the model with only first class constraints the condition for the 
generating function of the Dirac transformation
to be the physically equivalent one was given
in \cite{hori_6}.
The Cawley model\cite{cawley_1,cawley_2} is one of the models
in which Dirac's conjecture does not hold.
In order to generalize the formalism of \cite{hori_6} to the models with second class
constraints,
we present here a model with such constraint,
where Dirac's conjecture does not hold.
A new aspect which is absent in the model
with only first class constraints is
that the arbitrary functions appeared
in the general solution to the defining
equations of momenta are restricted.

%%%%%%%%%%%%%%%%%%%%%%%%%%%%%%%%%%%%%%%%%%%%%
\section{Cawley model}
%%%%%%%%%%%%%%%%%%%%%%%%%%%%%%%%%%%%%%%%%%%%%
%%%%%%%%%%%%%%%%%%%%%%%%%%%%%%%%%%%%%%%%%%%%%

In order to illustrate the canonical formalism
of \cite{hori_6},
let us start with the Cawley model\cite{cawley_1,cawley_2}.
The action of the Cawley model\cite{cawley_1} is
\be
%%%%%%%%%%%%%%%%%%%%%%%%%%%%%%%%%%%%%%%%%%%%
S_{\rm Cawley} = \int d\t L,
\sii
L
=
u^1u^2 + \fr12 q^3(q^2)^2,
%%%%%%%%%%%%%%%%%%%%%%%%%%%%%%%%%%%%%%%%%%%%
\ee
where $q^A,(A=1,2,3)$ 
are the coordinate variables
and $u^A,(A=1,2,3)$ are the corresponding
velocity variables.
The action is invariant under the
transformation
\be
%%%%%%%%%%%%%%%%%%%%%%%%%%%%%%%%%%%%%%%%%%%%
\d q^A = \d^A_1\e 
+ 
\d^A_3\fr{d}{d\t}
\left(
\fr{\dot{\e}}{q^2}
\right)
\sii
\d u^A = \fr{d}{d\t}(\d q^A),
\sii (A=1,2,3)
%%%%%%%%%%%%%%%%%%%%%%%%%%%%%%%%%%%%%%%%%%%%
\eq{caw_GTR}
\ee
where $\e$ is an arbitrary infinitesimal
parameter.

The Euler-Lagrange equations obtained by varying $q^A$ are
\be
%%%%%%%%%%%%%%%%%%%%%%%%%%%%%%%%%%%%%%%%%%%%
{\rm [EL]_1}=  \dot{u}^2 =0,
\sii
{\rm [EL]_2}= \dot{u}^1 - q^3q^2=0,
\sii
{\rm [EL]_3}= q^2=0.
%%%%%%%%%%%%%%%%%%%%%%%%%%%%%%%%%%%%%%%%%%%%
\eq{Ca_EL}
\ee
Since the lagrangian does not contain $u^3$,
the time development of $u^3$(and $q^3$) is
not determined, so we put $\dot{u}^3=c$,
where $c$ is an arbitrary function of $q$'s
and $u$'s.
We regard $q^3$ as an unphysical variables.
The consistency of the Euler-Lagrange equations
requires 
\be
%%%%%%%%%%%%%%%%%%%%%%%%%%%%%%%%%%%%%%%%%%%%%
\ell_1\deff q^2=0,
\sii
\ell_2\deff u^2=0,
%%%%%%%%%%%%%%%%%%%%%%%%%%%%%%%%%%%%%%%%%%%%%
\ee
which are called lagrangian constraints\cite{kami_1},
and should be satisfied in the initial condition
for the differential equations (\ref{Ca_EL}).
Equation $\ell_1=0$ is the third eqation of (\ref{Ca_EL}) and is called first order 
lagrangian constraint, while $\ell_2=0$ 
is obtained by time derivative of $\ell_1=0$
and is called second order one.

Let us proceed to the canonical theory
according to \cite{hori_6}.
Denote the canonical conjugate of $q^A$ as
$\p_A$.
Since the Hessian matrix, $M_{AB}=\del W_B/\del u^A$, where $W_B\deff \del L/\del u^B$,
is singular,
the defining equations of momenta,
$\p_A=W_A(q,u)$,
have not unique solution for the velocity
variables.
Hence we consider the {\it general solution} to
the equation and
denote it as $u^A=\hat{U}^A(q,\p)$.
The following functions on the coordinate-velocity
space play an impotant role:
\be
U^A_{\rm pb}(q,u)\deff \hat{U}^A(q,W(q,u)).
\ee
We call a function $A(q,u)$ the pull-back of
a function $\hat{A}(q,\p)$ if
$A(q,u)=\hat{A}(q,W(q,u))$,
and denote $A=\hat{A}_{\rm PB}$.
For example , $U^A_{\rm pb}(q,u)$ are the pull-back
of $\hat{U}^A(q,\p)$.
The functions $\hat{U}$'s contain
some arbitrary functions, which
play the similar role as the
Lagrange multipliers in the Dirac theory.
In the present model they are
$
%%%%%%%%%%%%%%%%%%%%%%%%%%%%%%%%%%%%%%%%%%%%%
\hat{U}^1 = \p_2,
~
\hat{U}^2 = \p_1,
~
\hat{U}^3 = \hat{v}(q,\p),
%%%%%%%%%%%%%%%%%%%%%%%%%%%%%%%%%%%%%%%%%%%%%
$
where $\hat{v}$ is an arbitrary function
of the canonical variables.

The hamiltonian in the present
formalism is defined by
\be
%%%%%%%%%%%%%%%%%%%%%%%%%%%%%%%%%%%%%%%%%%%%%
H
=
\p_A\hat{U}^A - L(q,\hat{U}),
%%%%%%%%%%%%%%%%%%%%%%%%%%%%%%%%%%%%%%%%%%%%%
\ee
and nothing is added\cite{hori_6}.
In the usual approach the hamiltonian
is defined by using $\dot{q}$'s instead of
$\hat{U}$'s.
But the meaning of $\dot{q}$'s
is obscure, since we are arguing on the phase space.
Kamimura\cite{kami_1} developed a
generalized canonical formalism
on the space spanned by  $(q,\dot{q},\p)$.
He introduced the concept of generalized
canonical quantity which plays a role of
canonical variables in the usual formalism.

Now the primary cinstraint  is $\vf \deff \p_3 = 0$.
Hamiltonian of the present model is
expressed as
\be
%%%%%%%%%%%%%%%%%%%%%%%%%%%%%%%%%%%%%%%%%%%%%
H = v\vf + q^3\c_1 + \p_2\c_2 ,
\sii
\c_1 \deff -\fr12 (q^2)^2,
\sii
\c_2 \deff \p_1.
%%%%%%%%%%%%%%%%%%%%%%%%%%%%%%%%%%%%%%%%%%%%%
\ee
The first order and the second order 
secondary constraints are
$\c_1=0$ and $\c_2=0$, respectively.
From the definition of the
hamiltonian we see
\be
%%%%%%%%%%%%%%%%%%%%%%%%%%%%%%%%%%%%%%%%%%%%%
\hat{U}^{A}
&=&
\pdel{H}{\p_{A}}
,
\sii
{\rm mod}~\vf,
\eq{0_U=dH/dp}
%%%%%%%%%%%%%%%%%%%%%%%%%%%%%%%%%%%%%%%%%%%%%
\ee
since we have $\p=W(q,\hat{U}(q,\p))$ on
the constrained sub-space.
An orbit O in the velocity-coordinate space 
is mapped by $\F: (q,u) \mapsto (q,\p=W(q,u))$ to an orbit $\hat{\rm O}$ in the 
phase space.
Since $\p=W(q,\hat{U}(q,\p))$ on $\hat{\rm O}$,
by differentiating it with respect to $\t$ along
the orbit
we have
\be
%%%%%%%%%%%%%%%%%%%%%%%%%%%%%%%%%%%%%%%%%%%%%
\dot{\p}_A
&=&
-\pdel{H}{q^A}
+
\left[
{\rm [EL]_A}
+
(\dot{q}^B - u^B)
\pdel{W_A}{q^B}
\right]_{u=\hat{U}},
\sii
{\rm mod}~\vf.
\eq{0_pdot=-dH/dq}
%%%%%%%%%%%%%%%%%%%%%%%%%%%%%%%%%%%%%%%%%%%%%
\ee
At this stage, however, 
the solution orbit
of the Euler-Lagrange equations in the
velocity-coordinate space has 
no counterpart in the phase space
which is obtained by a full set of
canonical equations of motion.
In the Dirac recipe they are obtained by
the canonical variational principle
which imposes the hamiltonian action, $\dot{q}^A\p_A - H_{\rm T}$,
to be stationary,
where $H_{\rm T}$ 
is the canonical hamiltonian plus 
the Lagrange multiplier terms
assuring the constraints.
In the present method 
we do not adopt the Dirac recipe, and
we have only the relations 
(\ref{0_U=dH/dp}) and
(\ref{0_pdot=-dH/dq}).

In order to get the canonical equations of
motion,
we need relations which express $\dot{q}^A$
in terms of canonical variables.
We determine the relation by imposing 
that the resulting canonical equations
are equivalent to the Euler-Lagrange equations.
The correct choice turns out to be
\be
\dot{q}^A = \hat{U}^A(q,\p).
\eq{0_dotq=hatU}
\ee
In fact from
eqs.(\ref{0_U=dH/dp})-(\ref{0_dotq=hatU})
we see that 
the canonical equations of motion,
$
\dot{q}^A=\del H/\del \p_A,~
\dot{\p}_A=-\del H/\del q^A
$,
are equivalent to
$
{\rm [EL]}_A(U_{\rm pb})=
\dot{q}^A - U^A_{\rm pb}=0
$.
These equations are equivalent to the
second order Euler-Lagrange equations obtained
by eliminating $u$-variables.
As for the secondary constraints,
the pull-back of the $k$-th order secondary constraints are
shown to be the  $k$-th order lagrangian constraints where
$u$'s are replaced by $U_{\rm pb}$'s \cite{hori_6}.
(The pull-back of the primary constraints are 
of cause identity.)
Thus the pull-backed theory from the canonical 
theory is
completely described by the original lagrangian
with the notational change $u\rarw U_{\rm pb}$.

 \vss

The canonical equation of motion
for a function $\hat{F}(q,\p)$ is  expressed as
\be
%%%%%%%%%%%%%%%%%%%%%%%%%%%%%%%%%%%%%%%%%%%%%
\fr{d}{d\t}\hat{F}  = \{\hat{F},H\},
\ee
where the Poisson bracket is defined by
\be
\{\hat{F},\hat{G}\}
\deff
\pdel{\hat{F}}{q^A}
\pdel{\hat{G}}{\p_A}
-
\pdel{\hat{F}}{\p_A}
\pdel{\hat{G}}{q^A}
.
%%%%%%%%%%%%%%%%%%%%%%%%%%%%%%%%%%%%%%%%%%%%%
\ee

 \vss

Now let us examine Dirac's conjecture.
The transformation generated by
a linear combination of first class
constraints with arbitrary parameters
is called Dirac transformation\cite{hori_6}.
Dirac's conjecture claims that
Dirac transformation map 
a state to its physically equivalent one.
The all of the 
Poisson brackets among $\vf,\c_1$ and $\c_2$
vanish, and they constitute first class 
constraints.
Hence the Dirac transformation is generated by
\be
Q = \hat{\e}^0\vf + \hat{\e}^i\c_i,
\sii ({\rm sum}~{\rm over}~i=1,2)
\ee
where $\hat{\e}^n,(n=0,1,2)$ are
arbitrary infinitesimal quantities, 
and the transformation is
\be
\d_Qq^A 
=
\{q^A,Q\}
=
 \d^A_1\hat{\e}^2 + \d^A_3\hat{\e}^0
+ \{q^A,\hat{\e}^1\}\c_1
+ \{q^A,\hat{\e}^2\}\c_2,
\ee
where 
$
\e^n(q,u)\deff \hat{\e}^n(q,W(q,u)),
(n=0,1,2)
$.
If one chooses parameters, $\hat{\e}$'s, which do not depend
on $\p$, then 
the pull-back transformation
is
\be
\d_{\rm D} q^A= \d^A_1\e^2
+ 
\d^A_3\e^0,
\sii \d u^A = \fr{d}{d\t}(\d q^A).
\ee
Under the above variations the 
Euler-Lagrange equation ${\rm [EL]}_2=0$
varies as
\be
\d_{\rm D}{\rm [EL]}_2 = \ddot{\e}^2 - \e^0 q^2,
\ee
which cannot vanish identically at
the point where Euler-Lagrange equations
and the lagrangian constraints hold.
This means the breakdown of
Dirac's conjecture.

 \vss

For a function $\hat{F}(q,\p,\t)$
let us denote
\be
%%%%%%%%%%%%%%%%%%%%%%%%%%%%%%%%%%%%%%%%%%%%
\hat{F}^\sim \deff 
\pdel{\hat{F}}{\t}
+
\{\hat{F},H\}.
%%%%%%%%%%%%%%%%%%%%%%%%%%%%%%%%%%%%%%%%%%%%
\ee
We can  prove\cite{hori_6} that
\be
%%%%%%%%%%%%%%%%%%%%%%%%%%%%%%%%%%%%%%%%%%%%
\dot{F}(q,U_{\rm pb})
=
\hat{F}^\sim(q,W(q,u))
+
{\rm [EL]}_A(U_{\rm pb})
\left[
\pdel{\hat{F}}{\p_A}
\right]_{\rm PB},
%%%%%%%%%%%%%%%%%%%%%%%%%%%%%%%%%%%%%%%%%%%%
\ee
where 
$
F(q,u,\t)=\hat{F}(q,W(q,u),\t)
$.
Let us define the function $E(\t)$ by
$
E(\t) \deff F(q_{\rm sol},\dot{q}_{\rm sol})
$,
where $q^A_{\rm sol}$
is the solution to ${\rm [EL]}_A=0$.
Then it satisfies
$
\dot{E}(\t)  
= 
\left.\hat{F}^\sim\right|_{\rm PB}
(q_{\rm sol},\dot{q}_{\rm sol})
$.
If the generating function $Q$ of the Dirac transformation satisfies
$
Q^\sim = 0 ~~ {\rm mod}~ \vf,
$
then the transformation is a gauge
transformation.
This can be proved by a general theorem\cite{hori_6}, {\it i.e.},
$
Q^\sim 
=
\d_{\rm D} L\Big|_{u=\hat{U}} + 
$ total derivative, 
where $\d_{\rm D}$ stands for the
pull-back of the Dirac transformation.
In the present model we have
\be
%%%%%%%%%%%%%%%%%%%%%%%%%%%%%%%%%%%%%%%%%%%
Q^\sim
=
(\hat{\e}^{1\sim} - \hat{\e})\c_1
 + 
\left(
\hat{\e}^{2\sim} -  q^2\hat{\e}^1
\right)\c_2
\sii {\rm mod}~\vf.
%%%%%%%%%%%%%%%%%%%%%%%%%%%%%%%%%%%%%%%%%%%
\ee
Hence the Dirac transformation is
gauge transformation if
$
\hat{\e} = \hat{\e}^{1\sim},
~
\hat{\e}^1= \hat{\e}^{2\sim}/q^2
$.
The pull-backed transformation is
eqs.(\ref{caw_GTR}) with 
$\e(\t)=\e^2(q_{\rm sol},\dot{q}_{\rm sol})$,
which keeps the action 
be invariant.
Since $\hat{\e}$
is an arbitrary function of $(q,\p)$,
the parameter $\e(\t)$
is sufficiently arbitrary.
In fact the action is invariant for
completely arbitrary $\e(\t)$.

%%%%%%%%%%%%%%%%%%%%%%%%%%%%%%%%%%%%%%%%%%%%%%
%%%%%%%%%%%%%%% hori model %%%%%%%%%%%%%%%%%%%
%%%%%%%%%%%%%%%%%%%%%%%%%%%%%%%%%%%%%%%%%%%%%%
\section{A model with second class constraints}
%%%%%%%%%%%%%%%%%%%%%%%%%%%%%%%%%%%%%%%%%%%%%%
%%%%%%%%%%%%%%%%%%%%%%%%%%%%%%%%%%%%%%%%%%%%%%
%%%%%%%%%%%%%%%%%%%%%%%%%%%%%%%%%%%%%%%%%%%%%%

The Cawley model contains only the first class
constraints.
Let us examine whether model with second class constraints has the similar property or it 
needs modification of the general theory\cite{hori_6},
where it assumed there is no second class
constraint.

Consider the lagrangian
\be
%%%%%%%%%%%%%%%%%%%%%%%%%%%%%%%%%%%%%%%%%%%%%%%
L
=
u^1u^2 
+ 
q^3
\left(q^2 - \fr12 q^3\right).
%%%%%%%%%%%%%%%%%%%%%%%%%%%%%%%%%%%%%%%%%%%%%%%
\ee
The action is invariant under the
following transformation
\be
%%%%%%%%%%%%%%%%%%%%%%%%%%%%%%%%%%%%%%%%%%%%%%%
\d q^A 
= 
\d^A_1\e 
- 
\d^A_3\fr{\dot{\e}u^2}{q^2-q^3},
%%%%%%%%%%%%%%%%%%%%%%%%%%%%%%%%%%%%%%%%%%%%%%%
\eq{hori_GTR}
\ee
where $\e$ is an arbitrary infinitesimal
quantity.
The Euler-Lagrange equations are
\be
{\rm [EL]}_1\deff \dot{u}^2 = 0,
\sii
{\rm [EL]}_2\deff\dot{u}^1 - q^3 =0,
\sii
{\rm [EL]}_3\deff -(q^2 - q^3) = 0.
\ee
Since the time development of $u^3$
is not determined by the Euler-Lagrange equations,
$q^3$ is regarded as unphysical variable, and we set $\dot{u}^3 = c$,
where $c$ is tentatively an arbitrary function of $(q,u)$'s.

The first and the second order lagrangian
constraints are
\be
%%%%%%%%%%%%%%%%%%%%%%%%%%%%%%%%%%%%%%%%%%%%%%%%
\ell_1 \deff q^2 - q^3 = 0,
\sii
\ell_2 \deff u^2 - u^3 = 0.%%%%%%%%%%%%%%%%%%%%%%%%%%%%%%%%%%%%%%%%%%%%%%%%
\ee
Instead of the third order lagrangian
constraint we have the condition $c=0$.
It is important to note that
the unphysical variable $q^3$ must be of
the form $q^3=a\t + b$, with constant $a$ and $b$,
though the action has the gauge degrees of
freedom.
Otherwise the Euler-Lagrange equations
have no solution.
In most of the gauge model this is not
the case, {\it i.e.}, the origin of the gauge symmetry is the arbitrariness  of 
the unphysical variables.
It would be thought that 
the above curiosity comes from the fact
that the gauge transformation is
singular at the point satisfying $\ell_1=0$.
However, in the Cawley model the unphysical
variable $q^3$ is completely arbitrary,
though the gauge transformation is
singular at $\ell_1=0$.
The above property in the present model 
is related to the existence
of second class constraint
as is shown in the canonical theory.

The Hessian matrix and $W=\del L/\del u$ are
\be
%%%%%%%%%%%%%%%%%%%%%%%%%%%%%%%%%%%%%%%%%%%%%%%%%%%%%
M_{ij}
=
\mxxxb
  0  &  1 & 0 \\
  1  &  0 & 0 \\
  0  &  0 & 0
\mxxxe,
\sii
W_1 = u^2,
\sii
W_2 = u^1,
\sii
W_3 = 0.
%%%%%%%%%%%%%%%%%%%%%%%%%%%%%%%%%%%%%%%%%%%%%%%%%%%%%
\ee

The primary constraint is $\vf\deff \p_3 =0$,
and the general solution for $u$'s to the
defining equation of momenta, $\p = W$,
is
\be
%%%%%%%%%%%%%%%%%%%%%%%%%%%%%%%%%%%%%%%%%%%%%%%%%%
\hat{U}^1(q,\p) = \p_2,
\sii
\hat{U}^2(q,\p) = \p_1,
\sii
\hat{U}^3(q,\p) = \hat{v}(q,\p),
%%%%%%%%%%%%%%%%%%%%%%%%%%%%%%%%%%%%%%%%%%%%%%%%%%
\ee
where $\hat{v}$ is an arbitrary function of the
canonical variables.
We see
\be
U^{1,2}_{\rm pb} = u^{1,2},
\sii
U^{3}_{\rm pb} = v(q,u),
\ee
where $v(q,u)\deff \hat{v}(q,W(q,u))$.

The
hamiltonian is
\be
%%%%%%%%%%%%%%%%%%%%%%%%%%%%%%%%%%%%%%%%%%%%%%%%%%
H
&=&
\hat{v}\vf
+
\p_2\p_1
-
q^3
\left(
q^2 - \fr12 q^3
\right)
\nn
&=&
H_0
+
\hat{v}\vf -   q^3\c_1 + \p_2\c_2,
\sii
H_0 \deff \hat{v}\p_2 - \fr12 (q^3)^2,
%%%%%%%%%%%%%%%%%%%%%%%%%%%%%%%%%%%%%%%%%%%%%%%%%%
\eq{hori_H}
\ee
where
\be
%%%%%%%%%%%%%%%%%%%%%%%%%%%%%%%%%%%%%%%%%%%%%%%%%%
\c_1 \deff 
q^2 - q^3,
\sii
\c_2 \deff \p_1 - \hat{v}.
%%%%%%%%%%%%%%%%%%%%%%%%%%%%%%%%%%%%%%%%%%%%%%%%%%
\ee
The first and the second order
secondary constraints are
$\c_1 = 0$ and $\c_2=0$, respectively.
There is no third order secondary
constraints, but we have the condition for $\hat{v}$:
\be
\hat{v}^\sim = 0
\sii {\rm mod}~ (\vf,\c_1,\c_2).
\eq{hori_v,H=0}
\ee
The pull-back of the secondary constraints,
$\c_1=\c_2=0$, are the lagrangian
constraints, $\ell_1(U_{\rm pb})=\ell_2(U_{\rm pb})=0$,
and that of the condition (\ref{hori_v,H=0}) 
corresponds to 
the condition $c=0$ in the lagrangian formalism.

Introducing the new constraints by
\be
\f_2 \deff
\c_2 - \{\c_1,\hat{v}\}\vf + \{\vf,\hat{v}\}\c_1
=0,
\ee
the Poisson brackets among
$(\f_0,\f_1,\f_2)=(\vf,\c_1,\f_2)$
are
\be
\{\f_n,\f_m\}
=
K_{nm}
\deff
\mxxxb  0 & 1 & 0\\
       -1 & 0 & 0\\
        0 & 0 & 0
\mxxxe
\sii {\rm mod}~(\f_0,\f_1),
\sii (n,m=0,1,2).
\ee
Hence we have two second class constraints,
$\f_0, \f_1$
and one first class constraint, $\f_2$.
The condition (\ref{hori_v,H=0})
comes from the fact that $\c_2$ contains $\hat{v}$,
which is a consequence of $\{\f_0,\f_1\}=1$.
In general the emergence of the condition for 
the arbitrary function in the
general solution to the defining equations
of momenta is a characteristic feature
in the model with second class constraints.

Since we have only one firat class
constraint, $\f_2=0$, 
the Dirac transformation is
generated by $Q=\hat{\e}\f_2$,
with arbitrary function $\hat{\e}(q,\p)$.
Let us obtain the condition for $\hat{\e}$ 
that the Dirac transformtion is gauge
transformation.
The criterion is $Q^\sim = 0~$ mod $\vf$.

\be
%%%%%%%%%%%%%%%%%%%%%%%%%%%%%%%%%%%%%%%%%%%%%%%
\hat{\e}^\sim  
= 
\hat{\e}
\left(
\pdel{\hat{v}}{q^3} + c_2
\right),
\sii
c_1
+
c_2\pdel{\hat{v}}{q^3}
+
\pdel{\hat{v}^\sim}{q^3}
= 0,
\sii {\rm mod}~\vf,
%%%%%%%%%%%%%%%%%%%%%%%%%%%%%%%%%%%%%%%%%%%%%%%
\eq{hori_CGTR}
\ee
where $c_i$ are defined by
\be
\hat{v}^\sim \deff c_1\c_1 + c_2\c_2
\sii {\rm mod}~\vf.
\ee

The pull-back of the Dirac transformation is
\be
%%%%%%%%%%%%%%%%%%%%%%%%%%%%%%%%%%%%%%%%%
\d_{\rm D} q^1
=
\pdel{}{u^2}
(\e F),
\sii
%%%%%%%%%%%%
\d_{\rm D} q^2
=
\pdel{}{u^1}
(\e F),
\sii
%%%%%%%%%%%%
\d_{\rm D} q^3
=
\e K
+ 
\e^3F,
%%%%%%%%%%%%%%%%%%%%%%%%%%%%%%%%%%%%%%%%%
\ee
where
\be
K \deff \pdel{\f_2}{\p_3}
\Big|_{\rm PB},
\sii
F \deff \f_2\Big|_{\rm PB}.
\ee

A choice for the arbitrary function, 
$\hat{v}$, partially determines not only
the canonical formalism but restricts
the lagrangian formalism.
Hence, if one choses one of the general sotutions,
then the pull-back of it restricts
the unphysical variables expressed in terms
of $U_{\rm pb}$ in the lagrangian
theory. This is illustrated below.

The condition (\ref{hori_v,H=0}) restricts 
the function form of $\hat{v}$,
and is written as
\be
\left\{
\hat{v},\p_1\p_2 - q^3
\left(q^2 - \fr12 q^3\right)
\right\} 
+ 
\{\hat{v},\p_3\}\hat{v} = 0
\sii {\rm mod}~(\vf,\c_1,\c_2)
\eq{hori_NLeq4v}
\ee
which is a non-linear differential 
equation for $\hat{v}$.
It seems difficult to obtain
the general solution to (\ref{hori_NLeq4v}), 
but we find some solutions:
\be 
%%%%%%%%%%%%%%%%%%%%%%%%%%%%%%%%%%%%%%%%%%%%%%%
\hat{v} = 0,  \si  \vf,\si \c_1,\si \p_1, \si \fr{(q^3)^2}{2\p_2}.
\eq{hori_v=}
%%%%%%%%%%%%%%%%%%%%%%%%%%%%%%%%%%%%%%%%%%%%%%%
\ee
For different  choices of $\hat{v}$,
the gauge structure and the Dirac transformations
become different.
In the case $\hat{v}=\p_1$
there is no first class constraint,
and the gauge symmetries of the
canonical formalism are absent.
Let us examine the remaining four cases separately.  \vss

\noi (1) $\hat{v}=0,~\vf,~\c_1$:

In these cases $\f_2 = \p_1,~K = 0,~F = u^2$,
and the pull-back of the Dirac transformation is
\be
%%%%%%%%%%%%%%%%%%%%%%%%%%%%%%%%%%%%%%%%%
\d_{\rm D} q^A 
= 
\d^A_1(\e + u^2\e^1)
+
\d^A_2u^2\e^2
+
\d^A_3\e^3 u^2,
%%%%%%%%%%%%%%%%%%%%%%%%%%%%%%%%%%%%%%%%%
\ee
where $\e^A = [\del\hat{\e}/\del\p_A]_{\rm PB}$.
For simplicity let us take $\e^1=\e^2=0$,
then
the Euler-Lagrange equation, ${\rm [EL]}_2=0$,
varies under the transformation as
\be
%%%%%%%%%%%%%%%%%%%%%%%%%%%%%%%%%%%%%%%%%
\d_{\rm D}{\rm [EL]}_2 = \ddot{\e} - \e^3u^2,
%%%%%%%%%%%%%%%%%%%%%%%%%%%%%%%%%%%%%%%%%
\ee
which cannot vanish for any
choice of the unphysical variable, $q^3$.
Hence Dirac's conjecture does not hold.

The second term in (\ref{hori_CGTR})
is satisfied in these cases,
and the condition for the Dirac 
transformation to be
gauge transformation becomes
$\dot{\e} = {\rm [EL]}_A\e^A$.
In fact 
lagrangian varies under the Dirac
transformation as 
$
\d_{\rm D}L
=u^2(\dot{\e} - {\rm [EL]}_A\e^A) + {\rm T.D.}
$.
If one choose the parameter
$\e(\t)= \e(q_{\rm sol},\dot{q}_{\rm sol})$
as in the case of the Cawley model,
then we see $\dot{\e}=0$.
This exhibits no gauge symmetry.
Instead,
we choose the parameter as
$\del \e/\del u^A=0$.
Here $\del \e/\del u^3=0$
is automatically satisfied since
$\del W_A/\del u^3=0$.
Then
the condition (\ref{hori_CGTR}) becomes
$
\dot{\e} = -(q^2 - q^3)\e^3,
$
and the gauge Dirac transformation is expressed as
eq.(\ref{hori_GTR}).

 \vss

\noi (2) $\hat{v}=(q^3)^2/2\p_2$:
 
In this case
we have 
$\hat{v}^\sim = 0~{\rm mod}~\vf$.
In eq.(\ref{hori_H}) we see 
$H_0=0$,
so the hamiltonian itself is a
linear combination of constraint functions.
In fact, by direct calculations we have
$
%%%%%%%%%%%%%%%%%%%%%%%%%%%%%%%%%%%%%%%%%
\f_2 = H/\p_2
%%%%%%%%%%%%%%%%%%%%%%%%%%%%%%%%%%%%%%%%%
$.
For simplicity
let us express the Dirac transformation
in the case of $\del \hat{\e}/\del \p_A=0$:
\be
%%%%%%%%%%%%%%%%%%%%%%%%%%%%%%%%%%%%%%%%%
\d_{\rm D} q^1 = \e,
\sii
\d_{\rm D} q^2 = \fr{\e q^3(2 q^2 - q^3)}{2(u^1)^2},
%%%%%%%%%%%%%%%%%%%%%%%%%%%%%%%%%%%%%%%%%
\sii
\d_{\rm D} q^3 
= 
\e\left(\fr{q^3}{u^1}\right)^2
+
\fr{\e^3 H_{\rm PB}}{u^1}.
%%%%%%%%%%%%%%%%%%%%%%%%%%%%%%%%%%%%%%%%%
\eq{hori_v2_Dirac}
\ee
The Euler-Lagrange equations, 
${\rm [EL]}_1={\rm [EL]}_2=0$,
vary as
\be
%%%%%%%%%%%%%%%%%%%%%%%%%%%%%%%%%%%%%%%%%
\d_{\rm D}{\rm [EL]}_1
= \fr{d^2}{d\t^2}(\d_{\rm D}q^2),
\sii
\d_{\rm D}{\rm [EL]}_2 = \ddot{\e} - \d_{\rm D}q^3,
%%%%%%%%%%%%%%%%%%%%%%%%%%%%%%%%%%%%%%%%%
\ee
which cannot vanish 
simultaneously 
for any
choice of the unphysical variable, $q^3$.
Hence Dirac's conjecture does not hold.

The condition for the Dirac transformation
to be gauge transformation is
\be
%%%%%%%%%%%%%%%%%%%%%%%%%%%%%%%%%%%%%%%%%
\dot{\e} 
-
{\rm [EL]}_A\e^A
=
\e\fr{q^3}{u^1}.
%%%%%%%%%%%%%%%%%%%%%%%%%%%%%%%%%%%%%%%%%
\ee
We set again $\e^{1,2}=0$, then
we have
$
\e^3 = 
(\e q^3 - \dot{\e}u^1)/u^1(q^2 - q^3),
$.
Substituting it into (\ref{hori_v2_Dirac}),
we have the gauge Dirac transformation
under which the action is invariant.

 \vss

Finally, let us comment on the relation of
the second class constraints and the Dirac
bracket.
The Poisson brackets among $\f_0,\f_1$
are written as
\be
%%%%%%%%%%%%%%%%%%%%%%%%%%%%%%%%%%%%%%%%%
\{\f_i,\f_j\} = A_{ij} \deff
\mxxb 0  & 1\\
      -1 & 0
\mxxe,
\sii (i,j=0,1).
%%%%%%%%%%%%%%%%%%%%%%%%%%%%%%%%%%%%%%%%%
\ee
Using $A_{ij}$, the hamiltonian is
expressed as
\be
%%%%%%%%%%%%%%%%%%%%%%%%%%%%%%%%%%%%%%%%%
H = H' - \f_i(A^{-1})^{ij}\{\f_j, H'\},
\sii
H' \deff
-\fr12(q^3)^2 + \p_2(\f_2 + \hat{v}).
%%%%%%%%%%%%%%%%%%%%%%%%%%%%%%%%%%%%%%%%%
\ee
In the above expression the squares of the
constraint functions are omitted,
which are no effect on the canonical
equations of motion,
$
\dot{\hat{F}} = \{\hat{F},H'\}_{\rm D},
$
here the Dirac bracket is defined by
\be
\{\hat{F},\hat{G}\}_{\rm D}
\deff
\{\hat{F},\hat{G}\}
-
\{\hat{F},\f_i\}(A^{-1})^{ij}\{\f_j, G\}.
\ee

%%%%%%%%%%%%%%%%%%%%%%%%%%%%%%%%%%%%%%%%%%%%%
%%%%%%%%%%%%%%%%%%%%%%%%%%%%%%%%%%%%%%%%%%%%%
%%%%%%%%%%%%%%%%%%%%%%%%%%%%%%%%%%%%%%%%%%%%%
%%%%%%%%%%%%%%%%%%%%%%%%%%%%%%%%%%%%%%%%%%%%%
%%%%%%%%%%%%%%%%%%%%%%%%%%%%%%%%%%%%%%%%%%%%%
\section{Concluding remarks}
%%%%%%%%%%%%%%%%%%%%%%%%%%%%%%%%%%%%%%%%%%%%%
%%%%%%%%%%%%%%%%%%%%%%%%%%%%%%%%%%%%%%%%%%%%%
%%%%%%%%%%%%%%%%%%%%%%%%%%%%%%%%%%%%%%%%%%%%%
%%%%%%%%%%%%%%%%%%%%%%%%%%%%%%%%%%%%%%%%%%%%%
%%%%%%%%%%%%%%%%%%%%%%%%%%%%%%%%%%%%%%%%%%%%%

The arbitrary functions, $\hat{U}$'s, emerging in the general solution to the defining equations
of momenta play a  role similar as
the Lagrange multipliers, $\l$'s, in the Dirac recipe,
though the ways of the appearance of them
are entirely different.
$\l$'s are independent variables, 
and satisfy linear algebraic equations.
The solution to the linear equations
is a sum of a special solution and
the general solution to a homogeneous
equation.
The former is a consequence of the
presence of the second class constraints
and the latter corresponds to the first class ones.
In the present approach
$\hat{U}$'s generally satisfy non-linear differential equations due to the
presence of the second class constraints.

The Dirac recipe may be easy to treat
compared with that of the present paper.
However, logical validity of the method of Lagrange multiplier is obscure.
In fact the method may contain
contradiction in a simple model\cite{frenkel}.
On the other hand the non-linear differential equations emerged in the presence  of
the second class constraints  may cause a technical difficulty in constructing 
a general theory with such constraints.

 \vss	
%%%%%%%%%%%%%%%%%%%%%%%%%%%%%%%%%%%%%%%
%%%%%%%%%%%%%%%%%%%%%%%%%%%%%%%%%%%%%%%
%%%%%%%%%%%%%%%%%%%%%%%%%%%%%%%%%%%%%%%
%%%%%%%%%%%%%%%%%%%%%%%%%%%%%%%%%%%%%%%
%%%%%%%%%%%%%%%%%%%%%%%%%%%%%%%%%%%%%%%
%%%%%%%%%%%%%%%%%%%%%%%%%%%%%%%%%%%%%%%
%%%%%%%%%%%%%%%%%%%%%%%%%%%%%%%%%%%%%%%
%%%%%%%%%%%%%%%%%%%%%%%%%%%%%%%%%%%%%%%
%%%%%%%%%%%%%%%%%%%%%%%%%%%%%%%%%%%%%%%
%%%%%%%%%%%%%%%%%%%%%%%%%%%%%%%%%%%%%%%

\section*{Acknowledgement}

\noi The author is grateful to Dr. Alexei Deriglazov
 for valuable discussions.

%%%%%%%%%%%%%%%%%%%%%%%%%%%%%%%%%%%%%%%%%
%%%%%%%%%%%%%%%%%%%%%%%%%%%%%%%%%%%%%%%%%
%%%%%%%%%%%%%%%%%%%%%%%%%%%%%%%%%%%%%%%%%
%%%%%%%%%%%%%%%%%%%%%%%%%%%%%%%%%%%%%%%%%
%%%%%%%%%%%%%%%%%%%%%%%%%%%%%%%%%%%%%%%%%
%%%%%%%%%%%%%%%%%%%%%%%%%%%%%%%%%%%%%%%%%
%%%%%%%%%%%%%%%%%%%%%%%%%%%%%%%%%%%%%%%%%
%%%%%%%%%%%%%%% refferences %%%%%%%%%%%%%
%%%%%%%%%%%%%%%%%%%%%%%%%%%%%%%%%%%%%%%%%
%%%%%%%%%%%%%%%%%%%%%%%%%%%%%%%%%%%%%%%%%
%%%%%%%%%%%%%%%%%%%%%%%%%%%%%%%%%%%%%%%%%
%%%%%%%%%%%%%%%%%%%%%%%%%%%%%%%%%%%%%%%%%
%%%%%%%%%%%%%%%%%%%%%%%%%%%%%%%%%%%%%%%%%
%%%%%%%%%%%%%%%%%%%%%%%%%%%%%%%%%%%%%%%%%

\end{document}